\documentstyle[psfig2]{mn}

\newif\ifAMStwofonts
\AMStwofontstrue

\ifoldfss
  \newcommand{\rmn}[1] {{\rm #1}}

  \ifCUPmtlplainloaded \else
    \NewTextAlphabet{textbfit} {cmbxti10} {}
    \NewTextAlphabet{textbfss} {cmssbx10} {}
    \NewMathAlphabet{mathbfit} {cmbxti10} {} 
    \NewMathAlphabet{mathbfss} {cmssbx10} {} 
  \fi
  \ifAMStwofonts
    \ifCUPmtlplainloaded \else
      \NewSymbolFont{upmath} {eurm10}
      \NewSymbolFont{AMSa} {msam10}
      \NewMathSymbol{\upi}     {0}{upmath}{19}
      \NewMathSymbol{\umu}     {0}{upmath}{16}
      \NewMathSymbol{\upartial}{0}{upmath}{40}
      \NewMathSymbol{\leqslant}{3}{AMSa}{36}
      \NewMathSymbol{\geqslant}{3}{AMSa}{3E}

      \let\leq=\leqslant 
       
    \fi
  \fi
\fi 

\ifnfssone
  \newmathalphabet{\mathit}
  \addtoversion{normal}{\mathit}{cmr}{m}{it}
  \addtoversion{bold}{\mathit}{cmr}{bx}{it}
  \newcommand{\rmn}[1] {\mathrm{#1}}

  \newmathalphabet{\mathbfit} 
  \addtoversion{normal}{\mathbfit}{cmr}{bx}{it}
  \addtoversion{bold}{\mathbfit}{cmr}{bx}{it}
  \newmathalphabet{\mathbfss} 
  \addtoversion{normal}{\mathbfss}{cmss}{bx}{n}
  \addtoversion{bold}{\mathbfss}{cmss}{bx}{n}
  \ifAMStwofonts
    \ifCUPmtlplainloaded \else
      \UseAMStwoboldmath
      \makeatletter
      \new@mathgroup\upmath@group
      \define@mathgroup\mv@normal\upmath@group{eur}{m}{n}
      \define@mathgroup\mv@bold\upmath@group{eur}{b}{n}
      \edef\UPM{\hexnumber\upmath@group}
      \new@mathgroup\amsa@group
      \define@mathgroup\mv@normal\amsa@group{msa}{m}{n}
      \define@mathgroup\mv@bold\amsa@group{msa}{m}{n}
      \edef\AMSa{\hexnumber\amsa@group}
      \makeatother
      \mathchardef\upi="0\UPM19
      \mathchardef\umu="0\UPM16
      \mathchardef\upartial="0\UPM40
      \mathchardef\leqslant="3\AMSa36
      \mathchardef\geqslant="3\AMSa3E

      \let\leq=\leqslant 

    \fi
  \fi
\fi 

\ifnfsstwo
  \newcommand{\rmn}[1] {\mathrm{#1}}

  \DeclareMathAlphabet{\mathbfit}{OT1}{cmr}{bx}{it}
  \SetMathAlphabet\mathbfit{bold}{OT1}{cmr}{bx}{it}
  \DeclareMathAlphabet{\mathbfss}{OT1}{cmss}{bx}{n}
  \SetMathAlphabet\mathbfss{bold}{OT1}{cmss}{bx}{n}
  \ifAMStwofonts
    \ifCUPmtlplainloaded \else
      \DeclareSymbolFont{UPM}{U}{eur}{m}{n}
      \SetSymbolFont{UPM}{bold}{U}{eur}{b}{n}
      \DeclareSymbolFont{AMSa}{U}{msa}{m}{n}
      \DeclareMathSymbol{\upi}{0}{UPM}{"19}
      \DeclareMathSymbol{\umu}{0}{UPM}{"16}
      \DeclareMathSymbol{\upartial}{0}{UPM}{"40}
      \DeclareMathSymbol{\leqslant}{3}{AMSa}{"36}
      \DeclareMathSymbol{\geqslant}{3}{AMSa}{"3E}

      \let\leq=\leqslant 

    \fi
  \fi
\fi 

\ifCUPmtlplainloaded \else
  \ifAMStwofonts \else 
    \def\upi{\pi}
    \def\umu{\mu}
    \def\upartial{\partial}
  \fi
\fi

\title{An infrared study of Centaurus A}
\author[J.J. Bryant \& R.W. Hunstead]{J.J. Bryant\thanks{jbryant@physics.usyd.edu.au.                                          
\newline Colour images for this paper can be found at http://www.physics.usyd.edu.au/$^{\sim}$jbryant/cenapub.html}, R.W. Hunstead\\
            School of Physics, The University of Sydney, NSW 2006, Australia.}
\date{Accepted..., Received....}

\pagerange{\pageref{firstpage}--12}
\pubyear{1998}

\begin{document}
\maketitle

\label{firstpage}

\begin{abstract}

We present {\it J}, {\it H} and {\it K}\,-band images and 0.9--2.5\,$\umu$m spectra of the
nuclear regions of Centaurus A obtained with IRIS on the AAT. While {\it K} band
has a point source coincident with the nucleus, at {\it H} and {\it J} we identify
diffuse structure extending to the NE as a possible ionisation cone.
By considering the NIR colours we
show that the point-like {\it K}\,-band emission originates not from the nucleus
itself, but from dust which has been shock heated by nuclear outflows.
A $J-K$ image reveals a band of high extinction across the nucleus lying
perpendicular to the radio jet axis, as suggested by previous authors
(Israel et al. 1990, Turner et al. 1992).  We model the detailed structure of this
extinction image with a circumnuclear torus of diameter $240\pm20\,$pc, 
thickness $75\pm4\,$pc, tilt $80\pm2\,$degrees to the
line of sight and with the torus and radio jet axes aligned. 

\end{abstract}


\section[]{INTRODUCTION}

Centaurus A (NGC5128, Cen A) is one of the closest active galaxies. Well
known as having a double radio lobe structure extending to
$6^{\circ}$, details in the radio jets have been mapped down to
milliarcsecs. While its radio power is relatively modest, Cen A,
at a distance of only 3\,Mpc (Tonry \& Schechter 1990, Ford et
al. 1989), offers a unique opportunity to investigate in detail the
smaller scale structures which surround and fuel the central power
source.

Optical observations of active galactic nuclei (AGNs) usually classify
them as Seyfert 1, Seyfert 2 or starburst, with the Seyfert types
linked by the unified model (Antonucci 1993) to be the same intrinsic
class of object but viewed at different angles to the line of sight.
It has been proposed that a similar picture applies to radio-loud AGNs
in which relativistic radio jets from the nucleus can extend in
bipolar outflows out to scales of $\sim 10^{6}$\,pc. The collimation
of these jets occurs within a scale $<1$\,pc. The working model used
to explain these phenomena consists of a hot accretion disk around a
supermassive black hole, with a broad emission line region (BLR) of a
few parsecs close to the nucleus and contained within a much larger
circumnuclear torus.  A narrow emission line region (NLR) extends to
tens to hundreds of parsecs.  In the Seyfert 1 model, the active
nucleus is viewed at a line of sight close to the radio jet axis, such
that the BLR, NLR and active nucleus are observed directly.  Seyfert
2s, on the other hand, are aligned more edge-on such that the
circumnuclear torus obscures the direct view of the BLR and nucleus,
and the radio jets are close to the plane of the sky.

While starburst galaxies are powered by thermal rather than
non-thermal nuclear processes, it is becoming increasingly recognized
that there is a substantial overlap between starburst and Seyfert
galaxies, with many showing evidence of starburst regions surrounding
a Seyfert nucleus (Moorwood \& Oliva 1988).

Cen A is classified as a Seyfert 2 galaxy with a prominent
warped dust lane which totally obscures the nuclear regions at optical
wavelengths.  The dust lane includes a population of young stars, and
is considered to be the result of a merger $\sim10^{9}$ years ago
(Malin, Quinn \& Graham 1983) which has led to a range of phenomena
spanning most of the observable spectrum.  Because of lower dust
extinction, near IR ($\sim1$--$2.5\,\umu$m) and longer wavelength
radiation penetrates the foreground dust lane, giving a view into the
circumnuclear torus and NLR. At $2.2\,\umu$m there is a point source 
coincident with the nucleus (Giles 1986), but the emission becomes
extended along the radio jet axis (p.a. = $51^{\circ}$) at shorter
infrared wavelengths. Joy et al. (1991) described the
$1.3\,\umu$m feature as the {\it J}\,-band `jet'. Previous authors have found a
region of higher extinction on scales of $\sim10$--16 arcsec around
the nucleus; this has been attributed to a circumnuclear torus, but
the exact orientation has been unclear.  As expected, no broad lines have
been found, consistent with the BLR being fully obscured.
Polarisation at p.a.\,$\approx115^{\circ}$, parallel to the dust lane,
was measured by Bailey et al. (1986) and Packham et al.
(1996), along with a patch of polarisation just on the
nucleus and perpendicular to the radio jet axis at
p.a.\,$\approx145^{\circ}$.  

Detailed radio studies show an FRI (Fanaroff \& Riley 1974) morphology
with radio structure ranging from lobes extending to 300\,kpc, down to
an unresolved point source of only 0.01\,pc in the nucleus (Schreier,
Burns \& Feigelson 1981; Kellerman, Zensus \& Cohen 1997). Jets at
X-ray wavelengths follow the same axis as the radio jets, and the
nucleus is variable in both radio and X-rays (Morini, Anselmo \&
Moltini 1989). Turner et al. (1992) have reported variability in the
infrared, but this remains to be confirmed.

We present an analysis of {\it J} ($1.25\,\umu$m), {\it H} ($1.65\,\umu$m) and {\it Kn}
($2.2\,\umu$m) images, and 0.9--$2.5\,\umu$m spectra, of the nucleus
of Cen A, directed at revealing the structure and processes in
the nuclear region. Section 2 describes the observations and
reduction. Images and spectra are presented in Section 3, and Section
4 discusses the observational results in the light of a model for the
structure of the circumnuclear regions and the contribution of
starburst processes to the infrared power.

\section[]{OBSERVATIONS AND REDUCTION}

\subsection{Images}

\begin{table}
\caption{Log of Observations.}
\begin{tabular}{@{}lccccc}
UT Date & Obs. & $\lambda$  & Seeing & Photo- &  Integ.  \\
     & Type &  & (arcsec) & metric & Time (s)\\
1995 May 23 & spectrum & HK & 1.5 & no & 1200 \\
1995 May 23 & spectrum & IJ & 1.5 & no & 1800\\
1995 Aug 12 & image & Kn & 0.9 & yes & 840\\
1995 Aug 12 & image & H & 1.0 & no & 840\\
1995 Aug 13 & image & J & 1.0 & no & 840\\
\end{tabular}
\label{logObs}
\end{table}

Images of the nuclear regions of Cen A in the {\it J}, {\it H} 
and {\it Kn} bands
were obtained with IRIS (Allen 1992) at the f/36 cassegrain focus of
the Anglo-Australian Telescope (AAT), Siding Spring Observatory on the
nights of 12 and 13 August 1995 (see Table~\ref{logObs}). The
$128\times128$\,pixel HgCdTe array was operated with 0.24 arcsec
pixels on the sky.  Each image was comprised of a five-point mosaic
pattern, centred on the {\it Kn} point source and with offsets of between 4
and 5 arcsec for the surrounding frames.  Two extra frames were taken
at radial offsets of 21.2 and 35.4 arcsec to the south-west in each
case, to image a bright star in the field, which we used to register
the three final images. Standard stars were observed at each
wavelength immediately before and after each mosaic pattern. Sky
frames were offset on average by 300 arcsec from the nucleus.

Frames were flat fielded using dome flats, then sky subtracted in {\sc
figaro}.  The reduction routines are IRIS-specific tasks in {\sc
figaro}. Once cleaned with IRISCLEAN, a foreground star on the dust
lane was used to register the frames when mosaicing with IRISMOS.
Magnitude calibration, using the standard star observations, corrected
all three mosaics, although the {\it H} and {\it J} observations were not made in
photometric conditions.  The {\it Kn}\,-band observation was made in
photometric conditions and agreed to $\sim0.1$--$0.2$\,mag with Giles
(1986) {\it K}\,-band ($2.2\,\umu$m) photometry. The non-photometric {\it J}\,- and
{\it H}\,-band mosaics were then calibrated by scaling to match Giles' data.

Aperture photometry results are given in Table~\ref{ApPhot}.
Measurements were made using the {\sc iraf} RADPROF program, with
apertures centred on the {\it K}\,-band point source position and diameters chosen
to match those used by Turner et al. (1992). Their photometry was centred
at the $3.26\,\umu$m point source position which is coincident with
the {\it K} point source. The aperture diameters were 2.52, 3.57
and 5.04 arcsec.  The photometric {\it Kn}\,-band images are within
0.2\,mag of Turner's results in all three apertures, and the scaled {\it J}
and {\it H} bands are also within that tolerance.  As a background aperture
is required for photometry and any such aperture will lie on the
galaxy, an unavoidable systematic error is introduced; based on the
measured background counts we estimate that it could be 0.1--0.2\,mag.

\begin{table}
\caption{Aperture photometry centred on the {\it Kn} point source. Errors in each 
magnitude are $\leq\,0.2$\,mag.}
\begin{tabular}{@{}cccccc}
Aperture & J & H  &  Kn &  J$-$H & H$-$Kn \\
Diameter &   &    &     &        &        \\
( arcsec)     & (mag) & (mag) & (mag) & (mag) & (mag)\\
2.52 & 11.8 & 10.8 & 9.7 & 1.0 & 1.1\\
3.57 & 11.2 & 10.1 & 9.2 & 1.1 & 0.9\\
5.04 & 10.6 & 9.6 & 8.7 & 1.0 & 0.9\\
\end{tabular}
\label{ApPhot}
\end{table}

\subsubsection{Alignment of images}
\label{align}

The {\it J}, {\it H} and {\it Kn} images were aligned relative to each other to an
accuracy of $\sim0.1$ arcsec by fitting several bright stars in the
mosaiced field (outside the regions shown in this paper) with {\sc
iraf} RADPROF, then shifting the images to align the stars to a
fraction of a pixel. The nucleus position was also measured with
RADPROF by fitting the {\it Kn}\,-band point source and noting its location to
$\sim0.1$\,pixel. As the {\it K} point source is
coincident with the nucleus defined by the radio position, it will be
referred to as the {\it K} nucleus. The relationship between the real
nucleus and the {\it K} nucleus will be examined in section~\ref{NatIR}. 

To put coordinates on the images, the FITS header parameters were set
to define the pixel scale of the image (0.24 arcsec pixel$^{-1}$) and
the coordinates of the nucleus pixel position.  We adopted the most
accurate published position for the nucleus of RA\,=\,13 25 27.46,
Dec\,=\,$-$43 01 10.2 (J2000), with uncertainties of $\pm1$ arcsec
(Schreier et al. 1998). The images are taken to be exactly north
upwards and East to the left. Therefore, while the absolute
coordinates on our images are accurate to $\pm1$ arcsec, the
registration of our overlaid images is accurate to $\sim0.1$ arcsec.
The coordinates on the model images discussed later were set in the
same way, with the nucleus being defined by the centre of symmetry of
the model.

\subsection{Spectra}

{\it IJ} and {\it HK} echelle spectra were obtained with IRIS on the AAT on 23 May
1995 (see Table~\ref{logObs}) at a spectral resolution of 400; the
overall wavelength coverage was from 0.9--$2.5\,\umu$m. The 13-arcsec
slit was oriented east-west, with a spatial scale of 0.79 arcsec
pixel$^{-1}$. With the {\it K}\,-band nucleus positioned 3.5 arcsec either
side of the centre of the slit in alternate frames, the subtraction of
pairs of frames removed sky and background galaxy.  Spectra were also
formed from the slit regions that were subtracted as background
galaxy, to test the uniformity of the background removal in the two
positions. The resulting spectra were very similar.  Although, at 7
arcsec from the nucleus the subtracted galaxy spectrum may not be
representative of the background at the nucleus, spectra extracted
closer to the nucleus showed strong variations with slit position and
were therefore unsuitable (see section~\ref{offnucresults}).


Each echelle frame was flat fielded and cleaned in {\sc figaro}.
Frames were then coadded and spectra extracted from the central 4
pixels.  IRISFLUX was used to calibrate the spectra from observations
of the standard star bs4903 (Allen \& Cragg 1983).  As the conditions
were not photometric, the resulting spectra only have relative flux
values. Argon lamp spectra were used for wavelength calibration and
wavelengths were corrected to a heliocentric frame.

\begin{figure*}
\centerline{\psfig{figure=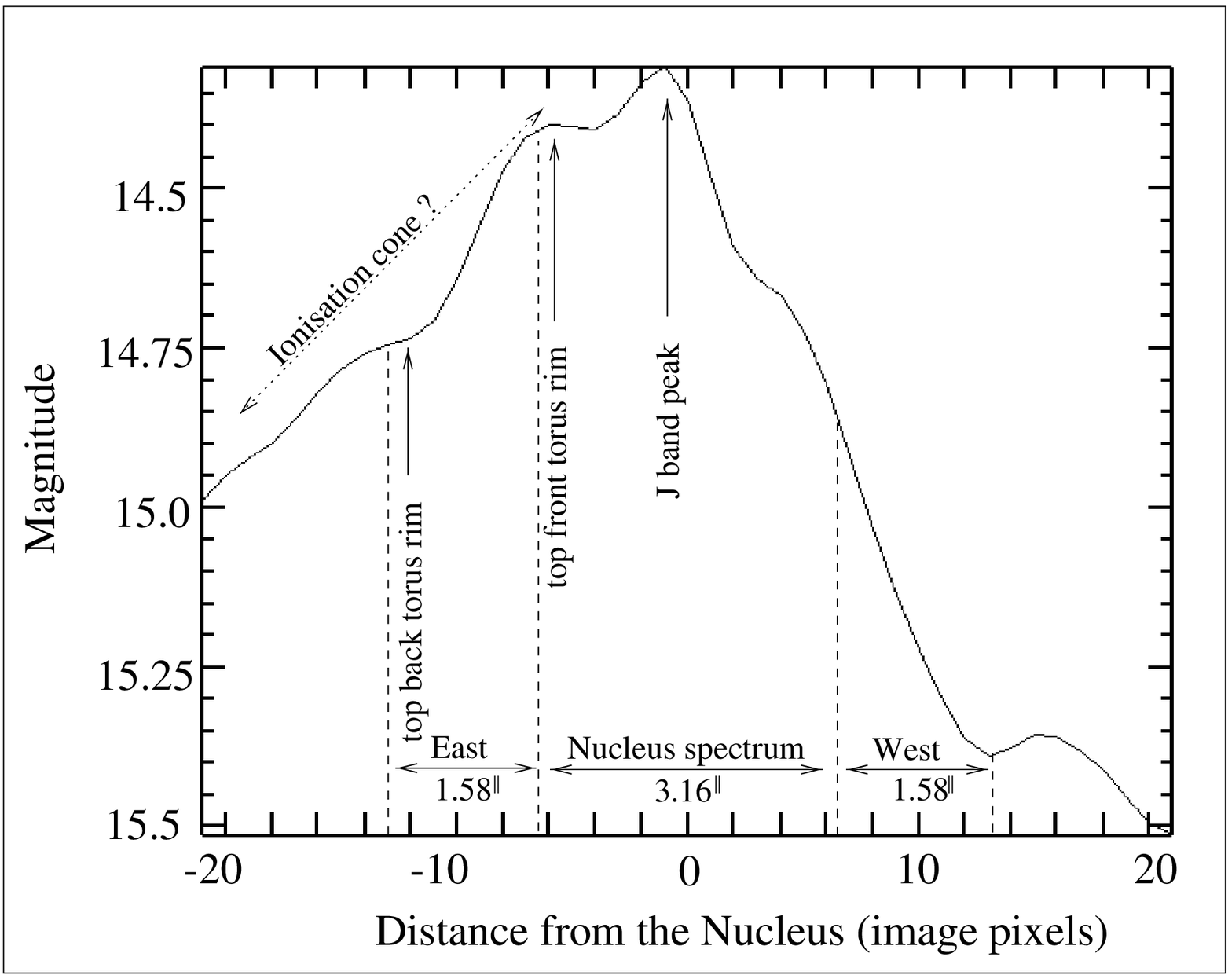,height=8cm}}
\caption{
An E-W cross section through the {\it J} image.  The regions corresponding
to the nuclear, east and west spectra are marked. The torus and
ionisation cone models are discussed in sections~\protect\ref{physparam}
and~\protect\ref{Jbandjet} respectively.}
\label{slitprofile}
\end{figure*}

As the intensity profile along the slit in each band clearly showed
structure away from the central peak, we extracted two additional
regions from positions 2.37 arcsec (3 pixels) E and W of the nuclear
region. These east and west regions only spanned 2 spatial pixels and
therefore have a lower S/N than the nuclear spectrum. In
Figure~\ref{slitprofile} the regions extracted to form the nuclear,
east and west spectra are marked on an E-W cross section through the
{\it J}\,-band image (see also Figure~\ref{slit}). Reduction procedures for
the east and west regions were the same as for the nucleus.  As
discussed in section~\ref{offnucresults}, the east and west spectra
exhibit very different characteristics.

\section[]{RESULTS}

\subsection{Images}

The {\it J}, {\it H} and {\it Kn}\,-band images are shown in Figure~\ref{HJKimages}. Only
in the {\it Kn}\,-band do we see a point source, while at {\it J} and {\it H} the diffuse
peaks are offset from the {\it K} nuclear position, with a significant
extension of the shorter wavelength emission to the north-east.

\begin{figure}
\caption{
Mosaic images of Cen A in {\it J}, {\it H} and {\it Kn} bands. The arrow marks a
foreground star. Each image is $38\times38$ arcsec.}
\label{HJKimages}
\end{figure}

\subsubsection{Nuclear size and intensity}

Previous estimates of the angular size of the nuclear source include
{\it K}\,-band FWHM $\leq0.5$ arcsec (Giles 1986), $3.26\,\umu$m FWHM $<1.5$
arcsec (Turner et al. 1992) and X-ray FWHM $<0.3$ arcsec (Feigelson et al. 1981).
The deconvolved FWHM of our {\it Kn} nucleus, measured from {\sc figaro}
CENTERS is $<0.7$ arcsec, or $<10$\,pc at a distance
of 3\,Mpc. The peak of the {\it J}\,-band emission is offset from the {\it Kn}
nuclear position by 0.4 arcsec.  As the {\it J}\,- and {\it H}\,-band peaks are
diffuse with no strong point source component, it can be assumed that the
underlying galaxy and circumnuclear emission dominates the {\it J} and {\it H}
bands. 


To measure the AGN point source contribution to the total {\it Kn}\,-band
emission, cuts were taken through the position of the {\it Kn} nucleus in
all three images, along a line of constant extinction (see
section~\ref{physparam}).  The position angle of this line, measured
from the extinction contour image, is $129\pm10^{\circ}$; Turner et al.
(1992) had previously reported 121.5$^{\circ}$. One-pixel-wide cuts
were made using PVECTOR in {\sc iraf}.  A point spread function (psf),
taken from a {\it Kn}\,-band foreground star 32 arcsec SW of the nucleus, was
then iteratively scaled and subtracted from the {\it Kn}\,-band profile until the
residue matched the shape of both the {\it H} and {\it J} profiles and, hence, of
the assumed underlying galaxy.  Photometry on the scaled psf gave the
measured point source flux as a percentage of the total {\it Kn}\,-band flux
as $39\pm6$ per cent in a 2.5 arcsec aperture.  Previous estimates of
the {\it K}\,-band point source contribution in the same aperture are 23 per
cent (Turner et al. 1992) and $30\pm5$ per cent (Packham et al. 1996).  

\subsubsection{Nuclear variability}

The nucleus of Cen A is known to be variable at radio wavelengths on
timescales of days (Kellerman 1974), and variations on timescales of
minutes to years have been recorded in X-rays (Morini et al. 1989). In
the mid-infrared, there is good evidence for variability: Lepine, Braz
\& Epchtein (1984) found an increase of a factor of 5 from 1971 to 1981
at $3.6\,\umu$m, while Turner et al. (1992)
reported a decrease of $\sim$2.5 times over $\sim$5 years (from 1983
to 1987) at $3.26\,\umu$m. The evidence for {\it K}\,-band ($2.2\,\umu$m)
variability, however, has been inconclusive, and at shorter NIR
wavelengths the underlying stellar contribution---presumed to be
constant---is far too dominant over the nuclear emission to detect
variability.  Furthermore, there is a foreground star 8.8 arcsec from
the nucleus (arrowed in Figure~\ref{HJKimages}) which appears to be
variable and is within some of the measurement apertures of the early
papers.

\begin{figure*}
\centerline{\psfig{figure=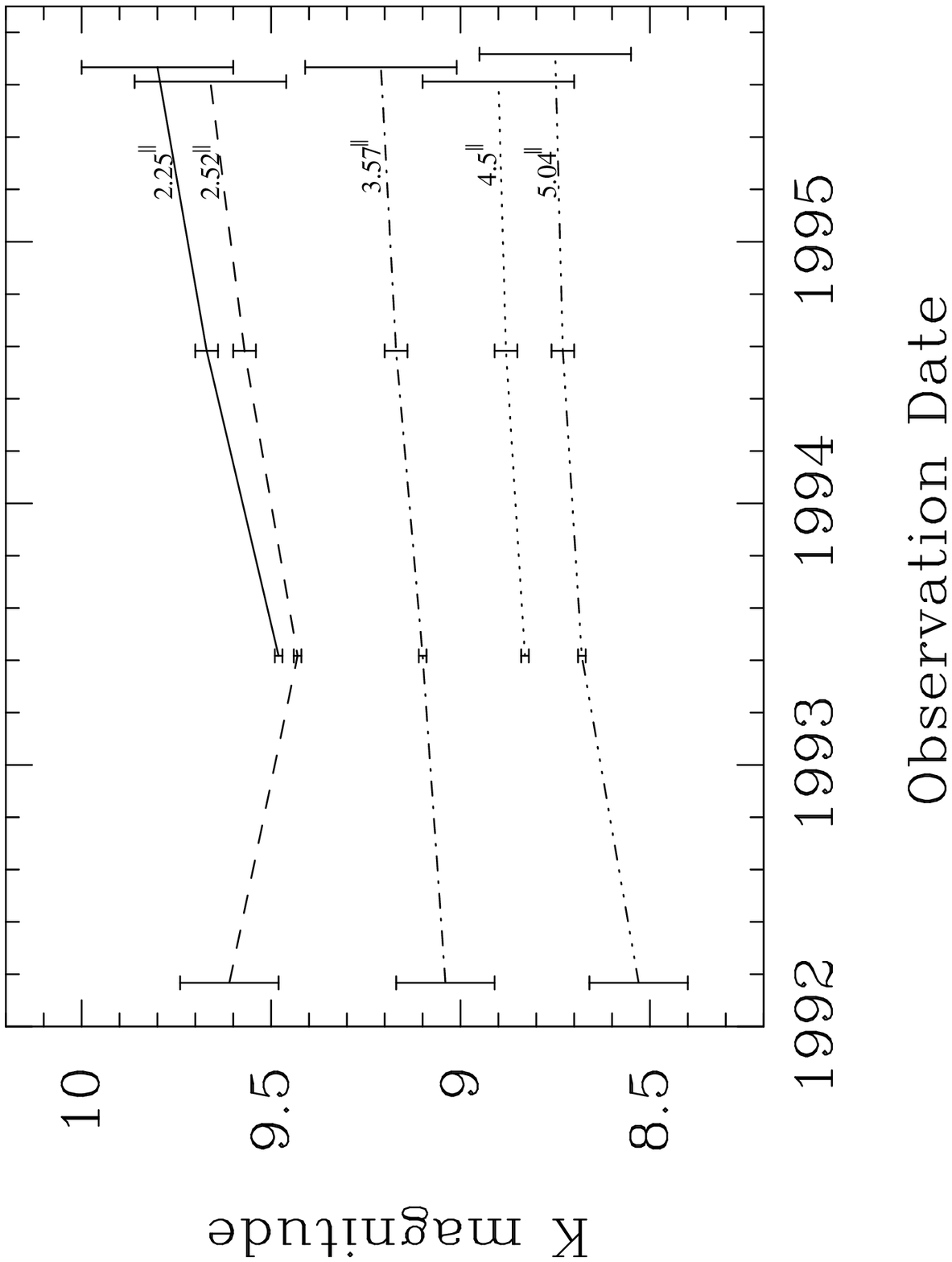,height=10cm,rotate=90}}
\caption{
{\it K}\,-band magnitude of the Cen A nucleus vs epoch of observation for a
range of photometric apertures. The data plotted are for Feb 1992
(Turner et al. 1992), May 1993 and Jul 1994 (Packham et al.
1996) and Aug 1995 (this paper).  Aperture diameters are marked.  Aug
1995 error bars are offset for clarity.}
\label{kvariab}
\end{figure*}

If {\it K}\,-band variability could be established, it would give an
independent measure of an AGN contribution to the {\it K}\,-band flux, as the
background starlight is presumably constant. To investigate {\it K}\,-band
variability, data taken at 4 epochs from Feb 1992 to Aug 1995 were
plotted in Figure~\ref{kvariab} for 5 different aperture sizes.  The
1993--1995 data were all taken with a {\it Kn} filter with a passband of
2.0--2.32\,$\umu$m, while the {\it K} filter for the 1992 data was centred
on 2.23\,$\umu$m. There is no consistent trend in variability with
aperture. The maximum change of $\sim0.3$\,mag occurs in the smallest
apertures.  Given that the point source contribution is $\sim$23--39 per
cent (from section 3.1.1), a dimming by 0.3 mag corresponds to a
decrease by a factor of 1.8--2.4 in the {\it K}\,-band point source.  However,
such a change is inconsistent with the constant or increasing
point-source contributions from 1993/1994 (Packham et al. 1996) to
Aug 1995 (this paper), suggesting that calibration errors may dominate
Figure~\ref{kvariab}.  We conclude that there is no evidence to date
for significant {\it K}\,-band variability and, therefore, that the {\it K}\,-band
point source emission is probably not coming directly from the AGN;
this issue is discussed further in section~\ref{NatIR}.

 
\subsection{Spectra}

\subsubsection{Nucleus}

The continuum shape of the nuclear spectrum suggests a residual
contribution from the surrounding galaxy as well as from the nucleus.
Although a galaxy component has been subtracted, it is very difficult
to remove the galaxy contribution entirely because of the patchy
nature of the dust near the nucleus.  This limitation is taken into
account when interpreting the emission line data.


In measuring line intensities, a local continuum was fitted and
subtracted from the regions around key lines, using Starlink {\sc
dipso}. The {\sc iraf} package SPLOT was used to measure wavelengths,
flux and FWHM for lines identified at the nucleus and the two adjacent
off-nuclear positions (see Figure~\ref{spec}).  Table~\ref{LinesNuc}
lists the emission line data for the three lines of sight. Quoted limits
were estimated from uncertainties in continuum placement and
extinction.  


\begin{table*}
\centering
\begin{minipage}{170mm}
\caption{Emission line data for the nucleus, east and west spectra. 
Flux densities have been corrected for a visual extinction A$_{\rmn
V}$ of 8.3 mag, as discussed in section 3.2.2 (see also
Table~\protect\ref{ExtincR}).}
\begin{tabular}{@{}lccccccccccc@{}}
 & & \multicolumn{4}{|c|}{Nucleus} &   & \multicolumn{2}{|c|}{$2.37$ arcsec
West} & & \multicolumn{2}{|c|}{$2.37$ arcsec East}\\
\cline{3-6} \cline{8-9} \cline{11-12}
Emission & & Dereddened & Upper & Deconv. & Upper &   & Dereddened & Upper & & Dereddened & Upper\\
Line & $\lambda$& Flux Density & limit & FWHM & limit &    & Flux Density & limit & & Flux Density & limit  \\
 & ($\umu$m) & \multicolumn{2}{|c|}{($10^{-17}$ W m$^{-2}$)} & \multicolumn{2}{|c|}{(km s$^{-1}$)} &   &  \multicolumn{2}{|c|}{($10^{-17}$ W m$^{-2}$)} & & \multicolumn{2}{|c|}{($10^{-17}$ W m$^{-2}$)}  \\
S\,{\sc iii} & 0.953 & 98 & 200 & 520 & 710 &   & 60 & 130 & & not detected & \\
He\,{\sc i} & 1.083 & 43 & 78   & 1000 & 1500 &   & 22 & 42   & & detected & \\
 Pa$\gamma$ & 1.094 & 19 & 34 & 1800 & 2400 &   & 7& 13 & & not detected & \\
 $[$Fe\,{\sc ii}$]$  & 1.257 & 12 & 18 & 600 & 700 &  & 8 & 13 & & 1.4 & 15 \\
Pa$\beta$ & 1.282 & 16 & 24 & 1580 & 1760 &   & 3 & 5 & & 0.4 & 4\\
$[$Fe\,{\sc ii}$]$ & 1.644 & 7 & 9 & 900 & 1200 &   & & detected\footnote{see text} &  & detected$^{\footnotesize a}$ & \\
H$_{2}(1-0)$S(3) & 1.958 & $<$4 &  & - & &  &  $<3$ &  & & $<$1 &    \\
H$_{2}(1-0)$S(2) & 2.034 & 3.1 & 3.8 & 520 & 730 &   & 1.4 & 1.6 & & not detected &  \\
He\,{\sc i} & 2.058 & $<$0.3 &  &  $<$830 & &   & not detected  &  & & not detected &  \\
He\,{\sc i} triplet & 2.112 & 2.4 & 2.8 & 470 & 720 &  &  0.9 & 1.0 & & not detected &  \\
H$_{2}(1-0)$S(1) & 2.122 & 5.1 & 6.0 & 530 & 700 &    & 1.6 & 1.9 & & inconclusive & \\
Br$\gamma$ & 2.166 & 2.7 & 3.2 & 590 & 890 &    &  0.6 & 0.7 & & 0.06 & 0.15 \\
H$_{2}(1-0)$S(0) & 2.223 & $<$1 &  & - &  &   & $<$0.5 &  & & detected &   \\
H$_{2}(2-1)$S(1) & 2.248 & $<$0.8 &  & - & &  &  $<$1.2 &  & & detected &   \\
H$_{2}(1-0)$Q(3) & 2.424 & 2.4 & 2.7 & - &  &   & 0.7 & 0.8 & & detected &   \\
\end{tabular}
\label{LinesNuc}
\end{minipage}
\end{table*}


\subsubsection{Extinction corrections}

 Amongst the detected lines, three pairs are commonly used to measure
the extinction to the line emitting region. Firstly, the [Fe\,{\sc
ii}]$\lambda\lambda\,1.257, 1.644\,\umu$m lines arise from the same
upper level and hence have an intensity ratio that is sensitive {\it
only\/} to reddening and not to temperature. Secondly, there is the
$\rmn{H}_{2}(1-0)\rmn{Q(3)}/\rmn{S(1)}$ ratio, which has large errors
due to the uncertainty in continuum placement around the Q-branch
lines.  Thirdly, the most commonly used line ratio is
Pa$\beta$/Br$\gamma$; these are two of the least obscured emission
lines and hence probe a larger depth. However, due to their relative
closeness in wavelength, a small error in the measurement of
Pa$\beta$/Br$\gamma$ can give a large error in A$_{\rm V}$. All three
line ratios were used to measure the extinction towards the nucleus
of Cen A, but the H$_{2}$ result was subsequently rejected
because of the large errors in the Q-branch lines.

The adopted extinction curve, $\tau\propto\lambda^{-1.85}$, is that of
Landini et al. (1984) and assumes that the extinction is purely foreground.
The adopted intrinsic line ratios are listed in Table~\ref{ExtincR}
along with the derived extinction values. A visual absorption
A$_{\rmn{V}}$ of 8.3\,mag (i.e., A$_{\rmn{V}} < 10.53$\,mag ) has been
adopted.  Although the A$_{\rmn{V}}$ derived from the [Fe\,{\sc ii}]
ratio agrees within error with that based on Pa$\beta$/Br$\gamma$, the
range is smaller in the latter.

\begin{table}
\caption{Intrinsic line ratios and calculated visual absorption 
towards the Cen A nucleus.}
\begin{tabular}{@{}ccc}
Line & Intrinsic& Calculated  \\
Ratio & Value &  A$_{\rmn{V}}$ \\
 & & \\
{\large $\frac{\rmn{H}_{2}(1-0)\rmn{S(1)}}{\rmn{H}_{2}(1-0)\rmn{Q(3)}}$} & 1.42 & $<$22 \\
 & & \\
{\large $\frac{\rmn{Pa}\beta}{\rmn{Br}\gamma}$} & 5.89 & $8.3^{+2.3}_{-5}$\\
 & & \\
{\large $\frac{\rmn{[Fe\,{\sc ii}]}\lambda\,1.257\,\umu
\rmn{m}}{\rmn{[Fe\,{\sc ii}]}\lambda\,1.644\,\umu \rmn{m}}$} & 1.386 &
$5.7^{+6.4}_{-3.5}$\\
 & & \\
\end{tabular}
\label{ExtincR}
\end{table}

Previous estimates of the extinction towards Cen A have ranged
widely, depending on the method used and the position in the galaxy.
Israel et al. (1990) corrected their line fluxes with an extinction
to the nucleus from
$\rmn {H}_{2}(1-0)\rmn{S(1)}/\rmn{Q(3)}\,=\,1.2^{+0.7}_{-0.4}$ of
A$_{\rmn{V}}=8.3$\,mag (coincidentally the same as our adopted value)
or, in his limiting case, A$_{\rmn{V}}<26$\,mag.  The corresponding
H$_{2}$ ratio from the present data gives A$_{\rmn{V}} < 22$\,mag.
While this is consistent with Israel et al., the larger errors
make it less reliable than Pa$\beta$/Br$\gamma$.


Measuring extinction to the infrared nucleus using silicate absorption
($9.7\,\umu$m), Israel et al. (1990) found A$_{\rmn{V}}$ to be
$15\pm4$\,mag, while others have measured values as high as
A$_{\rmn{V}}=22\pm5$\,mag (Becklin et al. 1971, Giles 
1986). We would expect the silicate absorption and X-ray
emission to probe to greater depth than the NIR 
lines and therefore show larger extinctions. On the other hand, the
peak of the optical emission, which is displaced from the infrared
nucleus by 5.8 arcsec, has an extinction of only a few magnitudes (Joy
et al. 1991, Giles 1986) as the optical
emission is foreground to most of the dust. 

The extinction-corrected line intensities are given in
Table~\ref{LinesNuc} and the corresponding ratios of useful line pairs
are listed in Table~\ref{LineRat}.

\begin{table}
\caption{Extinction corrected (A$_{\rmn{V}}=8.3$\,mag) emission 
line ratios for the nuclear spectrum.}
\begin{tabular}{@{}cc}
Ratio & Value \\ & \\
{\large $\frac{\rmn{H}_{2}(1-0)\rmn{Q(3)}}{\rmn{H}_{2}(1-0)\rmn{S(1)}}$} & $0.5^{+0.3}_{-0.2}$ \\
 &  \\
{\large $\frac{\rmn{H}_{2}(1-0)\rmn{Q(1-4)}}{\rmn{H}_{2}(1-0)\rmn{S(1)}}$} & $1.3^{+0.6}_{-0.6}$  \\
 &  \\
{\large $\frac{\rmn{H}_{2}(1-0)\rmn{S(2)}}{\rmn{H}_{2}(1-0)\rmn{S(1)}}$} & $0.6^{+0.4}_{-0.3}$ \\
 &  \\
{\large $\frac{\rmn{H}_{2}(1-0)\rmn{S(3)}}{\rmn{H}_{2}(1-0)\rmn{S(1)}}$} & $<$1.2  \\
 &  \\
{\large $\frac{\rmn{H}_{2}(2-1)\rmn{S(1)}}{\rmn{H}_{2}(1-0)\rmn{S(1)}}$} & $<$0.2 \\
 &  \\
{\large $\frac{\rmn{H}_{2}(1-0)\rmn{S(0)}}{\rmn{H}_{2}(1-0)\rmn{S(1)}}$} & $<$0.3  \\
 &  \\
{\large $\frac{\rmn{H}_{2}(1-0)\rmn{S(1)}}{\rmn{Br}\gamma}$} & $1.9^{+1.4}_{-0.8}$  \\
 &  \\
{\large $\frac{\rmn{H}_{2}(1-0)\rmn{S(1)}}{\rmn{[Fe\,{\sc ii}]}\lambda\,1.644}$} & $0.7^{+0.9}_{-0.3}$  \\
 &  \\
{\large $\frac{\rmn{[Fe\,{\sc ii}]}\lambda\,1.644}{\rmn{Br}\gamma}$} & $2.6^{+2.4}_{-1.4}$  \\
 &  \\
{\large $\frac{\rmn{[Fe\,{\sc ii}]}\lambda\,1.257}{\rmn{Pa}\beta}$} & $0.7^{+2.3}_{-0.5}$  \\
 &  \\
{\large $\frac{\rmn{Pa}\beta}{\rmn{Br}\gamma}$} & $6^{+7}_{-4}$ \\
 &  \\
\end{tabular}
\label{LineRat}
\end{table}

\subsubsection{Off-nuclear spectra}
\label{offnucresults}

Figure~\ref{spec} shows portions of the spectra containing the major
lines for the nucleus spectrum compared with the corresponding
off-nucleus spectra. The spectra centred 2.37 arcsec (3 spatial pixels
on the slit) east and west of the nuclear position are remarkably
different, with the west spectrum resembling the nucleus. In both the
east and west spectra, the [Fe\,{\sc ii}]$\lambda\,1.644\,\umu$m line
could not be measured accurately due to the lower signal-to-noise
ratio and the complication of it falling on overlapping echelle
orders.  

 
\begin{figure*}
\caption{To view this figure, check 
http://www.physics.usyd.edu.au/$^{\sim}$jbryant/figure4.html}
\label{spec}
\end{figure*}

The extinction to the west spectrum was measured from the
Pa$\beta$/Br$\gamma$ ratio to be A$_{\rmn{V}}=8.8$\,mag
(A$_{\rmn{V}}<22$\,mag), consistent with the nuclear value of
A$_{\rmn{V}}=8.3$. Uncertainties in the Pa$\beta$ line are responsible
for the large upper limit on A$_{\rmn{V}}$.  However, since we can
assume that the nucleus has the highest extinction, the upper limit on
the nuclear extinction was used to obtain upper limits for the west
line fluxes. The east spectrum has larger uncertainties in the line
strengths and, as Br$\gamma$ was only marginally detected, the
uncertainty in A$_{\rmn{V}}$ is large, with ${\overline
{\rmn{A}}_{\rmn{V}}}=0$\,mag and a limit of A$_{\rmn{V}}<11$\,mag.
Despite the large uncertainties, these results suggest there may be
less extinction to the east slit position than to the nucleus and
west positions. While these differences are not conclusive, they do
lend support to our model for the circumnuclear region discussed in
the following sections.

\section[]{ DISCUSSION}

\subsection{Physical parameters of the circumnuclear torus}
\label{physparam}

Israel et al. (1990) suggested that Cen A has a dense
circumnuclear torus with diameter $<$\,16 arcsec, or $<$\,230\,pc at a
distance of 3\,Mpc. They inferred the existence of the torus from the
very much higher extinction at the nucleus, and the size was
constrained by the non-detection of H$_{2}$ lines at positions 5
arcsec away from the nucleus in all directions except south. It has
been proposed that the H$_{2}$ `coats' the inner edge of the torus.
Turner et al.'s (1992) ROSAT observations showed a region of higher
extinction extending less than $\sim$\,14 arcsec and, along with
previous authors, noted the position angle (p.a.) of the higher
extinction, to be not at the $\sim115^{\circ}$ p.a. of the dust lane,
but at p.a.\,$\sim135$--$140^{\circ}$, perpendicular to the radio and
X- ray jet axis. This location constrains the ridge of extinction to
be associated with or feeding the accretion disk producing the jets,
and hence has been claimed as the circumnuclear torus. From
$^{12}$CO($2-1$) observations of the Cen A molecular gas,
Rydbeck et al. (1993) mapped the regions of highly red- and blue-shifted 
gas, which should correspond to the receding and approaching
edges of the torus. They concluded that the disk has an outer diameter
of between 9.6 and 16.6 arcsec, a rotational speed of 220 km s$^{-1}$,
a central velocity of 561 km s$^{-1}$ and a p.a. of $\sim145^{\circ}$.

An extinction map, made from the current data by subtracting
seeing-matched {\it J} and {\it Kn} mag arcsec$^{-2}$ images, is shown in
Figure~\ref{extincangles}.  The dust lane at
p.a.$\,=\,115\pm4^{\circ}$ shows up clearly, as does a region of
higher extinction across the nucleus at p.a.\,=\,$140\pm6^{\circ}$. We
believe that the latter feature can be identified with the
circumnuclear torus. The $J-H$ and $H-Kn$ colours at a selection of
points along this feature are all consistent with reddening of
E-galaxy light due to extinction.

Figure~\ref{torusa}(a) shows the central $\sim16\times16$ arcsec with
extinction contours overlaid on the greyscale extinction image.
North-east of the nucleus is a straight, well defined extinction
boundary $\sim2$ arcsec long at p.a.\,=\,140$^{\circ}$. We propose
that the band of high extinction SW of this boundary, across the
nucleus, defines the front of the putative torus.  The {\it J}\,-band `jet'
emission is unobscured to the NE of this boundary, in the region
marked X.  To the NW and SE of this boundary the extinction increases
abruptly into the regions marked Y, which we interpret as the back
inner edge of the torus.

In Figure~\ref{torusa}(b), we show the greyscale extinction image
overlaid with contours of the high velocity $^{12}$CO $J\,=\,2-1$
emission from Rydbeck et al. (1993).  The reference frame of
Rydbeck et al.'s image was shifted to align the nucleus with
our adopted position (see section~\ref{align}) to an accuracy of
$\pm\,0.5$ arcsec.  The highest velocity emission `patches' are
slightly north-east of a line through the nucleus at a p.a. of
140$^{\circ}$, lending support to the picture in which the torus is
tilted slightly towards the line of sight.

\subsubsection{Torus model}

On the basis of the interpretation in the preceding section, we now
propose a model for the size and orientation of the circumnuclear
torus.  While the model is by no means exclusive, it fits the observed
data if our interpretation of the extinction boundary NE of the
nucleus proves correct.  Figure~\ref{torusb}(a) \& (b) shows a cartoon
representation of our proposed model in which the circumnuclear torus
is tilted slightly to the line of sight, revealing the back inner edge
of the torus either side of the {\it J}\,-band emission `cone'.  In
Figure~\ref{torusb}(c) the model is overlaid with the extinction
contours to show how the observed features relate to the model.

\begin{figure*}
\caption{
The {\it J}\,-{\it K} extinction image for the central $30\times30$ arcsec showing
the dust lane at p.a.\,$\approx115^{\circ}$ and another region of high
extinction at p.a.\,$\approx140^{\circ}$, perpendicular to the radio
jet axis. The {\it K}\,-band nucleus is marked by the cross.}
\label{extincangles}
\end{figure*}

\begin{figure*}
\caption{
(a)Central $\sim16\times16$ arcsec of the {\it J}\,-{\it K} extinction image 
in grey scale and contours. 
The nucleus is marked by a white cross, and X and Y are marked as
discussed in the text as positions on a torus. The SE end of the proposed 
torus is concealed by the foreground dust lane which lies across 
the bottom of the image. Contour levels are 1.4, 1.45, 1.5, 1.55, 1.6,
1.65, 1.7, 1.75, 1.8, 1.85, 1.9, 2.0, 2.1, 2.2, 2.3, 2.4 mag
arcsec$^{-2}$. (b) Rydbeck et al. (1993) highest velocity 
$^{12}$CO $J\,=\,2-1$ contours overlaid 
on the extinction image. Contour levels span the range 24--40\,K km
s$^{-1}$ (steps 2.0\,K km s$^{-1}$). Both images are the same scale. The dotted
line is at p.a.$=140^{\circ}$}
\label{torusa}
\end{figure*}

\begin{figure*}
\caption{
(a) A cartoon representation of our model for the Cen A torus 
and ionisation cone.  (b) The
transparent model showing the hot dust region (20\,pc,
section~\protect\ref{NatIR}) as a sphere in the centre.  (c) The model with
the extinction image overlaid in contours. Contour levels are 1.4,
1.45, 1.5, 1.55, 1.6, 1.65, 1.7, 1.75, 1.8, 1.85, 1.9, 2.0, 2.1, 2.2,
2.3, 2.4 mag arcsec$^{-2}$. A white cross marks the nucleus.}
\label{torusb}
\end{figure*}

The constraints on the torus model imposed by the extinction image
were (i) projected height of the torus: $75\pm4$\,pc, (ii) 
projected distance from the nucleus to the front top rim: $24\pm2$\,pc,
and (iii) outer diameter measured from Figure~\ref{torusa}(b):
$240\pm20$\,pc. To meet these requirements the angle between the line
of sight and the torus axis, which is assumed to be parallel to the
radio axis, must be $80\pm2^{\circ}$. This is consistent with the
angle of 50--80$^{\circ}$ estimated from the radio jet-to-counterjet
surface brightness ratio (Tingay et al. 1998).

The model is not entirely constrained by the observations.  In
particular, the cross section of the torus and the inner torus
diameter are unknown. While an elliptical cross section was chosen for
convenience, we acknowledge that this is likely to be a gross
over-simplification. Israel et al. (1990) modelled the
possible torus configurations and, based on the constraints of the
high mass compared to column density, concluded that the torus must be
a `fat' disk with cavity size comparable to disk thickness. The inner
torus diameter was therefore chosen to be 80\,pc, although it could
vary significantly from this. Furthermore, the {\it J}\,-band feature
discussed later has an extension perpendicular to the jet which is
proposed (see section 4.3) to be emission from the back inner edge of
the torus and extends to the width of the opening in this model. The
only observational constraint on the torus inner diameter comes from
the width of the {\it J}\,-band feature as it appears just above the front
rim. As the {\it J} emission originates from inside the torus, this width of
38\,pc (2.5 arcsec) must be a lower limit on the torus inner
diameter.

\subsection{Nature of the infrared nuclear source}
\label{NatIR}

\begin{figure*}
\centerline{\psfig{figure=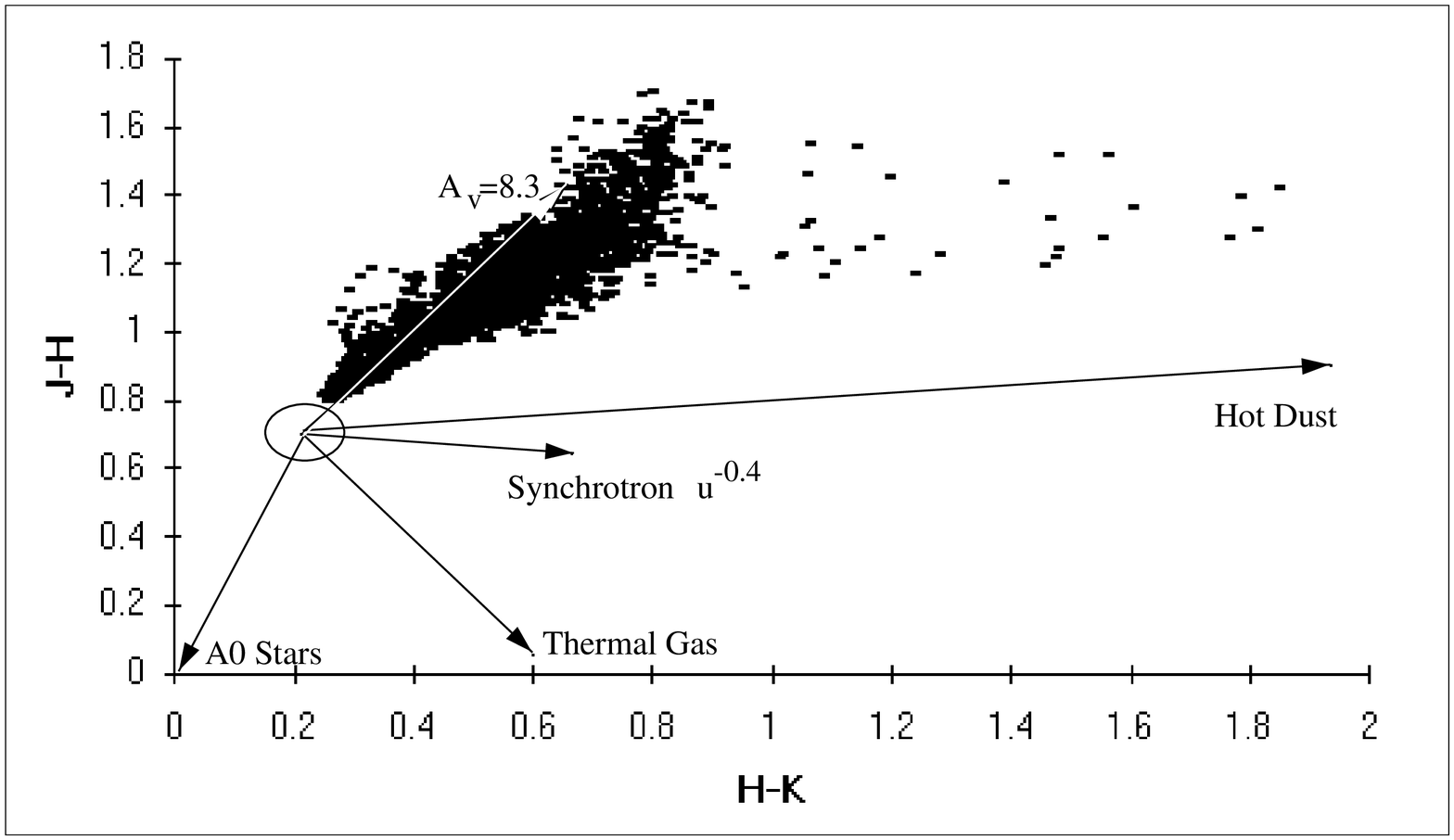,height=7cm}}
\caption{
$H-K$ vs. $J-H$ two-colour plot for pixels in an area of $12\times12$ 
arcsec centred on the nucleus. The A$_{\rmn{V}}$ vector 
represents 8.3 mag of extinction. The other vectors show
alternative processes, with the hot dust vector chosen for
$T\approx500$K. The ellipse shows typical unreddened E and S0 galaxy
colours.}
\label{N2col}
\end{figure*}

It is clear from the imaging data that the nucleus and surrounding
regions are heavily reddened. In order to determine if the reddening
is due solely to dust extinction, the two-colour plot in
Figure~\ref{N2col} was formed from values of the $50\times50$\,pixels
($12\times12$ arcsec) around the nucleus. 
The ellipse marks the typical colour range of unreddened E and S0
galaxies ($H-K=0.15$--0.25, $J-H=0.63$--0.73,
Frogel et al. 1978), with vectors showing the influence of
reddening from A$_{\rmn{V}}\,=\,8.3$ mag, hot dust ($T\sim500$\,K),
synchrotron and thermal gas emission, and post-starburst A0 stars.
The main bulk of the points lie along the extinction vector, which is
not surprising because of the dust lane and torus. When a larger pixel
area is included the distribution of points overlaps the elliptical
galaxy colours as we include the background galaxy contribution
further from the nucleus.

A notable feature of the two-colour plot is the collection of 27 points
that appear to lie along the extinction vector plus a hot dust vector.
This set of points comes from the central 1.4 arcsec covering the
{\it Kn}\,-band nucleus. As expected the nucleus is affected by the same dust
extinction as the rest of the circumnuclear regions but, in addition,
shows emission from hot dust from a diameter of only $\leq 20$\,pc. The dust
extinction to the nucleus determined from our spectral data was 8.3
mag. The vector of that length is clearly commensurate with the
extinction component to the nuclear reddening. For an extinction of
say 20 mag, the vector would have the same direction but extend three times
longer. Therefore, although the extinction estimate from the imaging
data remains uncertain, there is clear support for an extinction
of around 8 mag to the {\it Kn} emitting region, rather than the much larger
estimates in some previous papers (see section 3.2.2).

Figure~\ref{N2col} suggests that the observed emission at $2.2\,\umu$m 
and shorter wavelengths, is {\it not} coming directly from the nucleus.
Although the peak position at $2.2\,\umu$m agrees within errors with
the X-ray nucleus position (Turner et al. 1997), X-rays will
penetrate further than infrared radiation and hence should be
generated deeper within an object. X-ray extinction to the nucleus has
been estimated to be as high as A$_{\rmn{V}}\approx70$\,mag (Packham et
al. 1996, Morini et al. 1989), and the emission shows significant
variability.  The hot dust emission at {\it Kn}\,-band is therefore likely to
originate {\it not}\, from the nucleus, but from dust clouds close to
the nucleus which have been heated by the nuclear source. If this is
so, the size of the hot dust region (diameter\,$\approx20$\,pc) is
either a measure of the dimensions of the dust cloud, or the depth
into the dust cloud that the nuclear heating penetrates, and a region
of this size is shown around the nucleus of our model in
Figure~\ref{torusb}(b). Alternatively, the hot dust emission could
signal the presence of a circumnuclear starburst.

{\it K}\,-band polarimetry by both Bailey et al. (1986) and Packham 
et al. (1996) found polarisation perpendicular to the radio jet axis
at distances $<30$\,pc (2 arcsec) from the nucleus, with no
significant polarisation elsewhere in the nuclear region. Intrinsic
polarisation was ruled out by Packham et
al. (1996) in favour of a scattering model. Submillimetre
($800\,\umu$m and $1500\,\umu$m) polarimetry (Axon et al.
1994), where dust scattering is negligible, showed no ($<1$\%) nuclear
polarisation. Hence, rather than an intrinsically polarised source, the observed
polarisation is probably due to scattering.
The scattering clouds producing this
polarisation are likely to be close to the nucleus as the polarisation
angle is perpendicular to the radio jet axis and hence likely to be
influenced by the accretion processes fuelling the nucleus. In order
for the dust to survive so close to the nucleus, a toroidal geometry
around the radio axis may be required.  Therefore, the {\it Kn}\,-band source
is most likely hot dust emission from scattering clouds
$\leq 20$\,pc diameter covering the nuclear source.

\subsubsection{Excitation mechanisms for molecular hydrogen}

We now examine the origins of molecular hydrogen lines in the nuclear
spectrum.
Molecular hydrogen can be excited either by radiative fluorescence
through the absorption of soft-UV photons at low densities, or by
collisions in a warm gas.
Line ratios can be used to distinguish between these two alternatives.
For example, the ratios of $\rmn
{H}_{2}(2-1)\rmn{S(1)}$/$\rmn (1-0)\rmn{S(1)}$ and $\rmn
{H}_{2}(1-0)\rmn{S(0)}$/$\rmn {S(1)}$ should be
$\sim0.5$--$0.6$ for radiative UV fluorescence, but for collisional
excitation the first ratio is $<0.35$ and the second, 0.2--0.3. Our
measured line ratios in the nuclear spectrum are $<0.21$ and $<0.29$,
respectively, indicating that the molecular hydrogen in the nuclear spectrum
is probably collisionally excited. This result is predicted by studies of
populations of AGN and starburst galaxies by Fischer et al.
(1987) in which collisional excitation was the dominant excitation
mechanism.  Furthermore, Moorwood \& Oliva (1990) concluded that
radiative fluorescence at low densities is not an important mechanism
for exciting molecular hydrogen in AGNs.

The dominant heating mechanism for the collisional excitation of
H$_{2}$ can be derived from the line ratios in Table~\ref{HlineR}.
UV heating is due to
photons from the nucleus, incident on photodissociation regions, while
X-ray heating is responsible for exciting the molecular hydrogen on
the inner edge of a circumnuclear torus.
Hydrogen can also be
excited by shocks from nuclear driven winds incident on the NLR and
inner torus, in particular along the radio jet axis where the gas is
compressed by the radio-emitting outflow. The table lists the
predicted values for the diagnostic H$_{2}$ ratios, along with the
extinction corrected ratios measured from the current spectra. In all
cases, shock excitation is a key component, with a possible
contribution from X-ray heating.  Shock excitation of H$_{2}$ from
nuclear driven winds has been shown to be dominant in a number of
galaxies including NGC1068 (Kawara, Nishida \& Gregory 1990; Oliva and
Moorwood 1990). In addition, from a sample of Seyfert 2s, Veilleux,
Goodrich \& Hill (1997) found that shock excitation is the dominant
mechanism for exciting H$_{2}$, with X-ray heating also
playing a role.

\begin{table*}
\centering
\begin{minipage}{120mm}
\caption{Diagnostic ratios for collisional excitation of molecular hydrogen.}
\begin{tabular}{@{}cccccl@{}}
H$_{2}$ &UV & Shock & X-Ray & Measured for & Conclusion\\
Ratio      & Heating & Excitation & Heating &  Cen A nucleus & \\
$\frac{\rmn{H}_{2}(2-1)\rmn{S(1)}}{\rmn{H}_{2}(1-0)\rmn{S(1)}}$ & 0.5--0.6 & 0.1--0.2 & 0.01--0.02 & $<0.21$ & Shock or Shock+X-ray \\
 & & & & & \\
$\frac{\rmn{H}_{2}(1-0)\rmn{S(0)}}{\rmn{H}_{2}(1-0)\rmn{S(1)}}$ & 0.5--0.6 & 0.1--0.2 & 0.01--0.02 & $<0.29$ & Shock or Shock+X-ray\\
 & & & & & \\
$\frac{\rmn{H}_{2}(1-0)\rmn{Q(1-4)}}{\rmn{H}_{2}(1-0)\rmn{S(1)}}$ & $>2.6$ & $<2$ & & 1.3$^{+0.6}_{-0.6}$ & Shock\\
 & & & & & \\
$\frac{\rmn{H}_{2}(1-0)\rmn{S(1)}}{\rmn{Br}\gamma}$ & $<0.5$ & & &1.91$^{+1.4}_{-0.8}$ & Not UV\\
 & & & & & \\
$\frac{\rmn{H}_{2}(1-0)\rmn{S(3)}}{\rmn{H}_{2}(1-0)\rmn{S(1)}}$ & 0.5--0.7 & 0.8--1.1 & $\sim1.2$ & $<1.15$ & Shock or UV\\
\cline{2-4}
& \multicolumn{3}{|l|}{Kawara et al. (1990)} & &\\
& \multicolumn{3}{|l|}{Doyon et al. (1995)} & &\\
& \multicolumn{3}{|l|}{Veilleux et al. (1997)} & &\\
\end{tabular}
\label{HlineR}
\end{minipage}
\end{table*}

These results lend support to a model in which the H$_{2}$ lines are
generated on the inner edge of the circumnuclear torus by a
combination of shock and X-ray heating. The unified model for AGNs
(Antonucci 1993) predicts that in Seyfert 2s,
the H$_{2}$ lines should be partially obscured by the gas and dust in
the outer parts of the torus. The nuclear radiation does not penetrate
to the outer regions of the torus, so the temperature is not high
enough to excite H$_{2}$; instead the emission is reprocessed and
re-emitted in the FIR. In Seyfert 1s, the direct view of the inner
parts of the torus would be expected to show the excited molecular
hydrogen. Kawara et al. (1990) found H$_{2}$ emission more
strongly connected with Seyfert 1s than Seyfert 2s, as predicted by
this model. However, if in a Seyfert 2 galaxy the torus is tilted
slightly towards our line of sight, the H$_{2}$ from the inner back
edge of the torus should be observable over the front rim of the torus
(Alonso-Herrero, Ward \& Kotilainen  1996). Krolik \& Begelman (1988) claim to
have made such H$_{2}$ observations for several type 2 Seyferts. As
discussed below (and in Figure~\ref{torusb}, from section 4.1), the
Cen A circumnuclear torus appears to have a similar orientation,
revealing the H$_{2}$ from inside the torus.

\subsubsection{Excitation mechanisms for [Fe\,{\it II}] and
H\,{\it I} lines}

In AGNs, three main mechanisms have been put forward to explain
[Fe\,{\sc ii}] production. Firstly, starburst activity in or near the
nucleus can excite iron, either in the cooling tails of SNRs, in which
case there should be a correlation between the [Fe\,{\sc ii}] emission
and radio images, or through direct photoionisation by young OB stars
in a starburst region. Secondly, X-rays from the central source can
photoionise parts of the NLR clouds. Shocks induced by the interaction
of the radio jet with the ambient medium is the third option for iron
excitation, and predicts that the [Fe\,{\sc ii}] morphology should
match that of the radio jets.

If [Fe\,{\sc ii}] is to originate from either shock or photoionisation
processes in a circumnuclear starburst, it is first necessary to
consider whether Cen A has a composite Seyfert/starburst nucleus.
Several papers have found trends in emission line ratios from studies
of pure starbursts, composites and pure Seyfert nuclei. Moorwood and
Oliva (1988) found that the concurrent detection of H$_{2}$, [Fe\,{\sc
ii}] and H\,{\sc i} recombination lines was more probable in composite
nuclei than in pure starbursts, with none of their pure Seyferts
exhibiting all three species. They further found the ratio of
$\rmn {H}_{2} (1-0)\rmn{S(1)}/\rmn{Br}\gamma$ to be higher in
composites than starbursts.  [Fe\,{\sc
ii}]$\lambda\,1.644/\rmn{Br}\gamma$ ratios overlap between the two
groups, although Colina (1993) found this ratio must be
below 1.4 for pure starbursts. For Cen A, this ratio is 2.6 ($+2.4,
-1.4$), which makes pure starbursts unlikely, as expected.

The H$_{2}$ increase in composite nuclei reflects the additional
excitation of H$_{2}$ in the circumnuclear region by the AGN as
discussed in section 4.2.1. The mean value of $\rmn
{H}_{2}(1-0)\rmn{S(1)}/$[Fe\,{\sc ii}]$\lambda\,1.644$ from
Moorwood and Oliva's (1988) sample was $0.55\pm0.2$ for composites and
$0.34\pm0.15$ for H\,{\sc ii}/starbursts, with the current Cen A
spectra again clearly composite with a ratio of 0.73 ($-0.3,+0.9$).
Furthermore, a well documented correlation exists between [Fe\,{\sc
ii}]$\lambda\,1.257$ and H\,{\sc i} recombination lines from which
Puxley \& Brand (1994) found the mean values for the flux ratio
[Fe\,{\sc ii}]$\lambda\,1.257/\rmn{Pa}\beta$ to be $<0.18$
for starburst and $>0.35$ for AGN/composite galaxies respectively,
leaving Cen A clearly amongst the composites with a ratio of
0.74($+2.3,-0.5$).

We now explore the likely origin of [Fe\,{\sc ii}] in Cen A. 
Photoionisation
by hot stars is unlikely, because in photoionised gas iron is mostly
in a higher ionisation state than Fe$^{+}$ so the [Fe\,{\sc ii}] lines
are weak. Fast shocks, however, leave Fe singly ionised in their wake
from the destruction of Fe-rich interstellar grains. Therefore
[Fe\,{\sc ii}] is likely to originate from shocks either along the
radio axis or associated with SNRs in starbursts.

Several studies of starbursts and AGNs (Veilleux et al. 1997 and
refs therein) have demonstrated a relationship between [Fe\,{\sc ii}]
and radio emission in both classes of nucleus. Hence, any model of
[Fe\,{\sc ii}] generation may involve a process connected with the
synchrotron emission. This supports a picture in which the radio jets
induce fast shocks in the surrounding gas, with the line emission
resulting from the ionised post-shock gas. To uphold this radio
correlation in Cen A, a starburst shock origin would require an as yet
unsubstantiated association between the radio axis and the morphology
of the starburst region.

Based on the nuclear spectrum, the [Fe\,{\sc ii}] in Cen A is most
likely generated in shocks from nuclear outflow along the jet axis,
with the possibility of an additional starburst contribution.

\subsection{{\it J}\,-band `Jet' and nucleus}
\label{Jbandjet}

For consistency with the literature, the {\it J}\,-band feature is referred to
as a `jet' although the possible emission processes actually do not
allow for an outflow of infrared-emitting material from the nucleus. The `blue'
emission consists of a diffuse, approximately conical structure with
three positions of interest marked in Figure~\ref{Jjet}(a). The
brightest feature, at A, which we refer to as the {\it J}\,-band nucleus, is
located 0.4 arcsec east of the {\it Kn}\,-band nucleus. B marks a patch of {\it J}
emission stronger than the surrounding diffuse structure but weaker
than the {\it J} nucleus; it is located 2.2 arcsec NE of the {\it Kn}\,-band nucleus
and has some lateral extension. It is clear from
overlays with the extinction map (Figure~\ref{Jjet}(b)), that this
patch sits just above the front rim of our model torus, and hence
appears elongated perpendicular to the axis. B therefore marks the
start of the observable portion of the jet, which then extends to the
NE into region C.  Although B and C mark different portions of the
same feature, B has enhanced emission relative to C, as the view just
over the front rim of the torus includes a contribution 
from the inner edge of the far side of the torus.

\begin{figure*}
\caption{
(a) Circumnuclear region in {\it J}\,-band, where A marks the {\it J}\,-band `nucleus'
and B and C are along the ionisation cone. (b) The extinction image of
the torus in contours overlaid on the {\it J}\,-band image. Contours are 1.5,
1.55, 1.6, 1.65, 1.7, 1.75, 1.8, 1.85, 1.9, 2.0, 2.1, 2.2, 2.3\,mag
arcsec$^{-2}$. The cross marks the {\it K} nucleus.}
\label{Jjet}
\end{figure*}

\begin{figure*}
\centerline{\psfig{figure=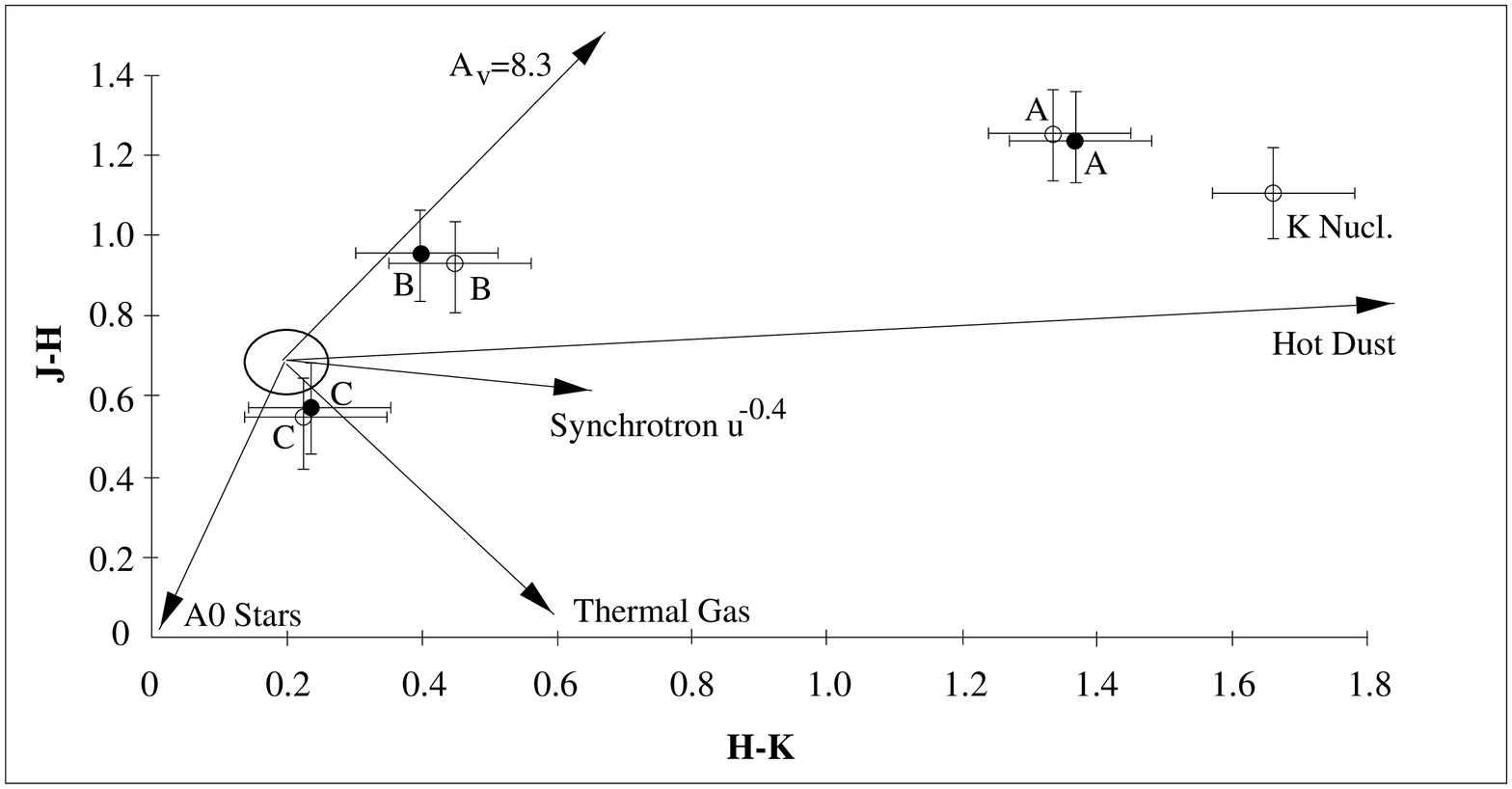,height=7cm}}
\caption{
two-colour plot for positions A, B and C on the blue jet, along with the
{\it Kn} nucleus colours. The closed and open circles represent aperture diameters
of 0.48 and 0.96 arcsec respectively.}
\label{J2col}
\end{figure*}

In order to explore the processes responsible for the `jet' emission, 
we need to
consider A, B and C separately. Aperture photometry was performed on
all three regions, with aperture diameters of 0.48 and 0.96
arcsec, chosen so the apertures did not overlap the adjacent regions.
The two-colour plot in Figure~\ref{J2col} maps these points along with
the previous aperture photometry for the {\it Kn}\,-band nucleus.  As in
Figure~\ref{N2col} typical
colours for E and S0 galaxies are marked by an ellipse and vectors
show the colours expected for each of the possible processes discussed
below. While the {\it J}\,-band nucleus has similar colours to the {\it Kn}
nucleus, as expected since it suffers the same extinction from the
torus and dust lane, the B and C positions have colours progressively
less affected by extinction and no significant hot dust component. As
discussed above, the hot dust cloud around the nucleus is less than
20\,pc (1.4 arcsec), and hence the {\it J} nucleus, offset from {\it K} by 0.4
arcsec, is within the same hot dust region as the {\it Kn} nucleus. 
The slight offset in position may be due to the
smaller depth into the torus seen at {\it J}. 

The `jet' in regions B and C, however, is clearly due to different processes.
Possible mechanisms for producing the jet include: (a) non-thermal
synchrotron emission, (b) scattered light from the AGN,
(c) starlight from blue stars, (d) thermal dust emission, or (e)
thermal gas emission (H\,{\sc ii}). From Figure~\ref{J2col} it is
clear that the colour of blue A0 stars eliminates (c) from being a
significant contributor to the jet emission.

Although the two-colour plot does not rule out a contribution from
synchrotron emission for B if it was in conjunction with dust extinction,
polarisation images by Bailey et al. (1986) and Packham et al.
(1996) show no
enhanced polarisation anywhere in the {\it J}\,-band `jet' region.
Furthermore, if the `jet' had a significant synchrotron component, we
would expect there to be corresponding radio emission. A $\lambda3.5$\,cm (1
arcsec resolution) ATCA radio image (Morganti 1998, private
communication) when overlaid on the {\it J}\,-band image, shows no correlation
between {\it J}\,-band and radio emission within $30\times30$ arcsec
around the nucleus.

Gas, dust or hot electrons located just above the torus have the
potential to scatter nuclear radiation towards the line of sight,
thereby revealing a hidden BLR (e.g. in NGC1068, Pogge \& de Robertis
1993).  However, the lack of polarisation in the Cen A `jet'
precludes scattering of nuclear radiation as a significant contribution.
This leaves the possibility of thermal emission from gas or
dust. We can speculate that the B position in the two-colour plot, could
be dust extinction plus a small hot dust emission component.
Alternatively, based on the C position, both B and C could include
thermal gas emission contributing around 10 percent of the {\it Kn}\,-band
flux. B has more extinction than C as it is closer to the highest
extinction regions of the foreground dust lane.

\subsubsection{Ionisation cone}
Hot dust emission at A and a cone morphology are strong indicators of
an ionisation cone, which traces ionising radiation from the nucleus
collimated along the radio jet axis, ionising the NLR gas which reradiates
in the infrared. The strongest circumstantial argument in favour of an
ionisation cone is the conical morphology of the `jet' emission.  Such
cones have been seen in several Seyfert 2 galaxies. Those with radio
jets and an obscuring torus, include NGC5728 (Wilson et al.
1993; Pogge 1989), NGC1365 (Phillips et al. 1983;
Storchi-Bergmann \& Bonatto 1991), NGC4388 (Pogge 1988; 1989; Corbin,
Baldwin \& Wilson 1988), and NGC5252 (Wilson and Tsvetanov 1994;
Tadhunter and Tsvetanov 1989). Two mechanisms can collimate the
radiation into a cone: firstly, the accretion disk and secondly,
shadowing from the molecular torus. In both cases the strong
correlation between the cone and radio jet axes seen in the objects
listed above (see Wilson and Tsvetanov 1994, Pogge 1989), is easily
explained. From the parameters of our simple model
(Figure~\ref{torusb}), the torus itself could not be collimating the cone.
However, the assumption of an elliptical cross section for the torus
is most likely an over-simplification and in fact the torus could be
thicker on the inside than the outside, which may collimate the cone
without affecting the correspondence of the model with the observed
torus. The cone opening angle adopted in the torus model was measured
from Figure~\ref{Jjet}(b) to be $60\pm10^{\circ}$, by extrapolating
back to the nuclear position. For comparison, Wilson and Tsvetanov's
(1994) sample of 11 ionisation cone galaxies showed a range of opening
angles between 40--92$^{\circ}$.

An ionisation cone should exhibit an emission line spectrum
characteristic of photoionisation by the nuclear source. Although the
current spectra do not include position C on the cone, a limited
spectrum is available at B. When the positions of the off-nuclear
spectra (see section 3.2.3 and Figure~\ref{slitprofile}) are overlaid
on the extinction and {\it J}\,-band images in Figure~\ref{slit}, the
differences in the east and west spectra become apparent as the west
spectrum sits on the front of the torus like the nuclear spectrum, but
the east position includes a contribution from the putative ionisation cone
near B.  Although there will be some contamination from the nucleus to
the East and West spectra, the differences are significant. It is not
surprising, therefore, that the east spectrum measures less extinction
(section 3.2.3), as the line emission is generated above the front rim
of the torus. Along the line of sight in the east spectrum position is
the {\it J} cone in the foreground with the inside back edge of the
circumnuclear torus behind. Therefore, not only should the east
spectrum have a contribution from ionisation lines along the radio jet axis as above, but
the excited molecular hydrogen lines, coating the inner torus, should
be included at higher extinction.
The west
spectrum has stronger [Fe\,{\sc ii}] and H$_{2}$ like the nucleus as
the west position includes scattered emission from the nucleus, inside
the torus.

\begin{figure*}
\caption{
{\it J}\,-band image overlaid with contours of the extinction image. Contours
are 1.6, 1.65, 1.7, 1.75, 1.8, 1.85, 1.9, 2.0, 2.1, 2.2, 2.3\,mag
arcsec$^{-2}$.  The rectangle marks the slit position.}
\label{slit}
\end{figure*}

Longer wavelengths penetrate further and therefore should be generated
deeper within an object. As the wavelength increases, the east
spectrum more closely resembles the nuclear and west spectra, showing
a strong correspondence with the nucleus for wavelengths longer than
$\sim2.1\,\umu$m. This trend is apparent in the line sections shown in
Figure~\ref{spec}. In the East spectrum, whilst undetected at shorter
wavelengths, H$_{2}$ lines are increasingly measurable at longer
wavelengths. However, even the shorter wavelength [Fe\,{\sc ii}] line
($1.257\,\umu$m) is detected, suggesting that [Fe\,{\sc ii}] is
generated from a shallower depth than the H$_{2}$ lines.  As the east
position is above the front rim of the torus and hence suffers less
extinction, if [Fe\,{\sc ii}] is excited along the radio jet axis the
H$_{2}$ may therefore originate on the inner back torus. 

The wavelength correspondence between the strong [Fe\,{\sc
ii}]$\lambda\,1.257\,\umu$m line and the {\it J}\,-band filter ($1.25\,\umu$m)
lends support to an ionisation cone that appears so strongly at
{\it J}\,-band. A fainter extension in the cone direction is also seen at
{\it H}\,-band (Figure~\ref{HJKimages}). The [Fe\,{\sc
ii}]$\lambda\,1.644\,\umu$m line lies in the {\it H} filter and was detected
but not measured (see section~\ref{offnucresults}).  The [Fe\,{\sc
ii}] lines do not confirm the ionisation cone, as they could be
shock excited by shocks
along the radio jet axis or shocks within small circumnuclear
starburst regions inside the torus area. However, based on the
morphology and colours of the {\it J} `jet', it is most likely an ionisation
cone of opening angle $60\pm10^{\circ}$, length $\sim100$\,pc (7
arcsec), extending NE from the nucleus along the radio jet axis. To
test this ionisation cone model, a map of spectra across the proposed
cone region and/or a narrow-band [Fe\,{\sc ii}] image should reveal the
distribution of the [Fe\,{\sc ii}] emission.

\section[]{CONCLUSION}
 
From {\it J}, {\it H} and {\it Kn}\,-band images and near-infrared spectra, a model of the
Cen A nuclear regions has been proposed consistent with the
observations. By no means an exclusive solution, the model depends on
the interpretation of the sudden change of extinction NE of the
nucleus as defining the top of a torus. This leaves a picture which
includes a point source nucleus at wavelengths longer than {\it K}\,-band,
from which collimated radio jets are produced. The nucleus is
partially obscured by a circumnuclear torus of outer diameter
$240\pm20$\,pc and thickness $75\pm4$\,pc which is tilted
$80\pm2^{\circ}$ from the line of sight, such that the inner back edge
of the torus is seen over the front rim. Scattering clouds surrounding
the nucleus are shock excited by nuclear outflows, which heat the
clouds to a depth of $\sim20$\,pc. {\it K}\,-band emission is generated in
this hot dust and not directly from the nucleus which is only seen at
longer wavelengths through more extinction.

We speculate that
radiation from the nucleus, collimated along the jet axis, produces a
cone of hot ionised NLR gas with opening angle $60\pm10^{\circ}$.
Reradiation from the cone, in the near-infrared, is preferentially at
shorter wavelengths and accompanied by [Fe\,{\sc ii}] emission lines.

\section*{ACKNOWLEDGMENTS}

We would like to thank Raffaella Morgani and Neil Killeen for
providing radio images prior to publication. We are also grateful to
Gustaf Rydbeck who supplied his CO image data for our
overlays. JJB acknowledges support from a University Postgraduate
Research Award from the University of Sydney.
RWH acknowledges support for observatory travel from an ARC Institutional
Grant.

\end{document}